\documentclass[12pt]{article}

\usepackage{amsfonts}

\newcommand{\sect}[1]{\setcounter{equation}{0}\section{#1}}
\newcommand{\subsect}[1]{\subsection{#1}}

\def\be{\begin{equation}}
\def\ee{\end{equation}}
\def\bea{\begin{eqnarray}}
\def\eea{\end{eqnarray}}

\def\k{\omega}

\def\om{J}
\def\j{J}
\def\m{M}
\def\bi{B}
\def\ei{I}
\def\e{E}

\def\sq{sq}

\def\Ke{{\mathbb K}}
\def\He{{\mathbb H}}
\def\Ze{{\mathbb Z}}
\def\Ree{{\mathbb R}}
\def\Cee{{\mathbb C}}

\def\Lam{{{\cal I}_\k}}

\def\a{\alpha}
\def\b{\beta}
\def\c{\gamma}
\def\ep{\varepsilon}

\def\cextiif{\alpha^{F}}
\def\cextiil{\alpha^{L}}
\def\cextiii{\beta}

\def\cexta{\alpha}
\def\cextb{\beta}
\def\cextg{\gamma}

\def\G{{\cal G}}

\def\sspp{U}

\begin{document}

\begin{center}
{\LARGE{\bf{The   families of
 orthogonal, unitary}}}
\medskip

{\LARGE{\bf{and quaternionic unitary Cayley--Klein}}}
\medskip

{\LARGE{\bf{algebras and their central extensions
\footnote{Contribution to the III International Workshop on {Classical
and Quantum Integrable Systems}. Edited by L.G. Mardoyan, G.S.
Pogosyan  and A.N. Sissakian. Joint Institute for Nuclear
Research, Dubna, pp. 58--67, (1998).}
}}}
\end{center}

\vskip0.5cm

\begin{center}
Francisco J. Herranz$^\dagger$ 
and Mariano Santander$^\ddagger$
\end{center}
\vskip0.3cm

\begin{center}
{\it $^\dagger$ Departamento de F\'{\i}sica, E.U. Polit\'ecnica,
Universidad de Burgos\\
E--09006 Burgos, Spain}
\end{center}

\begin{center}
{\it $^{\ddagger}$ Departamento de F\'{\i}sica Te\'orica,
Universidad de Valladolid \\
E--47011 Valladolid, Spain}
\end{center}

\vskip0.5cm

\begin{abstract}
The families of quasi-simple or Cayley--Klein 
algebras associated to antihermitian matrices over $\Ree$, $\Cee$
and $\He$ are described in a unified framework.
These three families include
simple and non-simple real Lie algebras which can be
obtained by contracting the pseudo-orthogonal algebras
$so(p,q)$ of the Cartan series $B_l$ and $D_l$, the special
pseudo-unitary algebras $su(p,q)$ in the series $A_l$, and
the quaternionic pseudo-unitary algebras $sq(p,q)$ in the series
$C_l$. This approach allows to study many properties for all
these Lie algebras simultaneously. In particular their
non-trivial   central extensions are completely
determined in arbitrary dimension. 
\end{abstract}

\vskip0.5cm


\sect{The three main families of CK algebras}

The aim of this contribution is twofold. First, we present the
structure of the three main series  of Cayley--Klein (CK)  algebras
(orthogonal, unitary and quaternionic unitary) as associated to
antihermitian matrices over
$\Ree$, $\Cee$ and $\He$ \cite{tesis,goslar}. These families include
real simple Lie algebras in the four Cartan series as well as many
non-simple Lie algebras which can be obtained from  the simple ones 
by a sequence of contractions \cite{graded,gradedb}.   The CK
approach is quite useful in the study of many structures 
associated to the algebras in a given CK family, 
such as their symmetric homogeneous spaces
\cite{dubna}, Casimir invariants \cite{casimir}, quantum deformations
\cite{qck}, etc. This formalism allows a
clear view of the behaviour of such properties under contraction.

And second, as an example of this approach, we focus on the
complete and general results concerning the non-trivial  central
extensions  \cite{ortogonal,unitario,symplectico}
for {\em all the algebras} in the three main  CK families.

We  consider $N$ real coefficients $\k=(\k_1,\k_2,\dots,\k_N)$ and 
 two-index coefficients  $\k_{ab}$ defined in terms of the former by
\be
\k_{ab}:=\k_{a+1}\k_{a+2}\cdots\k_b  \qquad
  a,b=0,1,\dots,N  \quad
  a<b  \qquad
\k_{aa}:=1 .
\label{aa}
\ee
Let $\Ke$ denote an associative division algebra; this can be either
the reals $\Ree$, complex
$\Cee$ or quaternions $\He$. The
space
$\Ke^{N+1}$ can be endowed  with a hermitian  (sesqui)linear
form $\langle \ .\ |\ . \ \rangle_\k :
\Ke^{N+1}\times \Ke^{N+1}\to  \Ke$ defined by
\be
\langle \>a|\>b \rangle_\k:=
\bar a^0 b^0 + \bar a^1 \k_1 b^1 + \bar a^2 \k_1 \k_2 b^2 + \dots
+ \bar a^N \k_1\cdots\k_N b^N =
\sum_{i=0}^N \bar a^i\k_{0i}  b^i 
\label{ab}
\ee
where $\>a,\>b\in \Ke^{N+1}$ and $\bar a^i$ means the  
conjugation in $\Ke$ of the component $a^i$. 
The underlying metric is   provided by the matrix
\be
\Lam = {\mbox{diag}}\ (1,\, \k_{01},\,\k_{02},\dots,\,\k_{0N}) =
       {\mbox{diag}}\ (1,\, \k_1,\,\k_1\k_2,\dots,\,\k_1\cdots\k_N) .
\label{ac}
\ee
The antihermitian CK family  over $\Ke$ is defined  as the family of
Lie algebras of the groups of linear isometries of this
hermitian metric over the space $\Ke^{N+1}$. Thus the isometry
condition for an element
$\sspp$ of the corresponding Lie group
\be
\langle \sspp \>a| \sspp \>b \rangle_\k= \langle \>a|  \>b \rangle_\k
\qquad \forall\, \>a,\>b\in \Ke^{N+1},
\label{ad}
\ee
leads to the following relations for the group element  $U$ and
generator $X$: 
\be
\sspp^\dagger\Lam \sspp=\Lam  \qquad X^\dagger\Lam +\Lam X=0  .
\label{ae}
\ee
For $\Ke=\Ree,\ \He$ these matrices span a simple Lie algebra. For 
$\Ke=\Cee$ the algebra so determined is not simple, but  becomes
simple if  we add the unimodularity condition:
\be
 \mbox{det}(U)=1  \qquad \mbox{trace}(X)=0 ,
\label{ag}
\ee
which is known as the {\em special} condition for the Lie 
algebra/group. Hence from condition (\ref{ae}) (and also (\ref{ag})
when 
$\Ke=\Cee$) we obtain  $\Lam$-antihermitian $(N+1)\times
(N+1)$ matrices over
$\Ke=\Ree,\Cee,\He$ which generate the   three  families of
quasi-simple or CK  algebras  displayed in the
following table \cite{tesis,goslar}:

\noindent
\begin{tabular}{llll}
$\Ke$&\  CK family&\quad Generators &\quad Dimension\\ 
\hline
$\Ree$&\ Orthogonal $so_{\k}(N+1)$ &\quad
$\om_{ab}=-\k_{ab}e_{ab}+e_{ba}$ &\quad $\frac 12 N(N+1)$\\[0.1cm]
$\Cee$&\  Complex special unitary &\quad
$\om_{ab}=-\k_{ab}e_{ab}+e_{ba}$ &\quad $(N+1)^2-1$\\ 
 &\quad $su_{\k}(N+1)$ &\quad
$\m_{ab}=i(\k_{ab}e_{ab}+e_{ba})$ &\quad  \\ 
 &\   &\quad
$\bi_l=i(e_{l-1,l-1}-e_{ll})$ &\quad \\[0.1cm]
$\He$&\  Quaternionic unitary &\quad
$\om_{ab}=-\k_{ab} e_{ab}+e_{ba}$ &\quad $\!2(N+1)^2
+(N+1)$\\ 
&\quad  $sq_{\k}(N+1)$ &\quad
$\m_{ab}^\a=i_a(\k_{ab}e_{ab}+e_{ba})$ &\quad  \\ 
 &\   &\quad
$\e_a^\a=i_\a  e_{aa}$ &\quad  \\ 
\hline
\end{tabular}

\noindent
Hereafter it is assumed that $a,b=0,1,\dots,N$ and $a<b$; 
$l=1,\dots,N$;    $i$ is the
complex unit;  $\a=1,2,3$ and $i_1=i$,  $i_2=j$,
$i_3=k$ are the usual quaternionic units;  and 
$e_{ab}$ is the $(N+1) \times (N+1)$ matrix with a single 1 entry in
row $a$, column $b$.

When all the coefficients $\k_a\ne 0$ these CK families include real
simple algebras in the four Cartan series:

\noindent
$\bullet$ The orthogonal CK family $so_{\k}(N+1)$ embraces the
pseudo-orthogonal algebras $so(p,q)$ with $p+q=N+1$ in the Cartan
series $B_{\frac N2}$ for even $N$, and $D_{\frac {N+1}{2}}$ for odd
$N$. 

\noindent
$\bullet$ The special unitary CK family $su_{\k}(N+1)$ comprises the
special pseudo-unitary algebras $su(p,q)$ with $p+q=N+1$ in the Cartan
series $A_N$.

\noindent
$\bullet$ The quaternionic unitary CK family $sq_{\k}(N+1)$ includes
the  quaternionic pseudo-unitary algebras $sq(p,q)$ with $p+q=N+1$
(the usual name for these algebras is $sp(p,q)$) in the Cartan
series $C_{N+1}$.

In all cases, $p$ and $q$ are the number of positive and  negative
terms in the matrix $\Lam$ (\ref{ac}), so when all $\k_a>0$ we
recover the compact real forms $so(N+1)$, $su(N+1)$ and $sq(N+1)$. If
we set one or several coefficients $\k_a$ equal to zero, we obtain a
non-simple Lie algebra. This process is equivalent to perform the
limit $\k_a\to 0$ which corresponds to carry out an In\"on\"u--Wigner
contraction
\cite{IW}. Therefore each CK family includes simple as
well as non-simple members and the latter appear by means  of
contractions. These non-simple Lie algebras   preserve properties
related to
 simplicity, and are also called quasi-simple 
\cite{Rozenfeld1}.
For instance, it has been shown in \cite{casimir}  that all the
algebras in the orthogonal CK family $so_{\k}(N+1)$ share the same
number of functionally independent Casimir operators no
matter how contracted is the Lie algebra. This not longer true if we
go beyond the CK family as the abelian algebra clearly shows: all its
generators are invariants.

These families of quasi-simple algebras can be obtained  by applying
the graded contraction theory \cite{Montigny,Moody} to any simple
member in  each  family (e.g., $so_{\k}(N+1)$ is provided by 
$\Ze_2^{\otimes N}$ graded contractions starting from $so(N+1)$
\cite{graded,gradedb}). An alternative approach to these
algebras can be found in \cite{Gromov}.

In the sequel we write down the Lie brackets for these three families
of Lie algebras. We remark that the values of each {\em real}
coefficient $\k_a$ can be reduced to $+1$,  $-1$ or 0 (by 
means of a rescaling), so that for a given $N$  each CK family
contains $3^N$ Lie algebras (some of them are isomorphic as abstract
Lie algebras).

The commutation relations of the  {\em orthogonal CK family} 
$so_\k(N+1)$ can be obtained from the matrix generators $J_{ab}$ and
read \cite{tesis,ortogonal}:
\be
[\om_{ab}, \om_{ac}] = \k_{ab} \om_{bc} \qquad
[\om_{ab}, \om_{bc}] = -\om_{ac} \qquad
[\om_{ac}, \om_{bc}] =  \k_{bc}\om_{ab}  \qquad  [\om_{ab},
\om_{de}] = 0 .
\label{ba}  
\ee
Hereafter whenever three indices $a,b,c$ appear they are always
assumed to verify $a<b<c$; whenever four indices
$a,b,d,e$ appear, $a<b$, $d<e$ and all of them   are assumed
to be different; and there is no any implied sum over repeated indices.

 The Lie brackets of the  {\em special unitary CK family} 
$su_\k(N+1)$ are
given by \cite{tesis,unitario}:
\be
\begin{array}{lll}
[\j_{ab},\j_{ac}] =\k_{ab}\j_{bc} &\qquad
[\j_{ab},\j_{bc}] =-\j_{ac} &\qquad
[\j_{ac},\j_{bc}] =\k_{bc}\j_{ab}\\{}
[\m_{ab},\m_{ac}] =\k_{ab}\j_{bc} &\qquad
[\m_{ab},\m_{bc}] =\j_{ac} &\qquad
[\m_{ac},\m_{bc}] =\k_{bc}\j_{ab} \\{}
[\j_{ab},\m_{ac}] =\k_{ab}\m_{bc} &\qquad
[\j_{ab},\m_{bc}] =-\m_{ac} &\qquad
[\j_{ac},\m_{bc}] =-\k_{bc}\m_{ab}\\{}
[\m_{ab},\j_{ac}] =-\k_{ab}\m_{bc} &\qquad
[\m_{ab},\j_{bc}] =-\m_{ac} &\qquad
[\m_{ac},\j_{bc}] =\k_{bc}\m_{ab} \\{}
[\j_{ab},\j_{de}]= 0 &\qquad
[\m_{ab},\m_{de}] =0 &\qquad
[\j_{ab},\m_{de}] =0 \\
\multicolumn{3}{c}{
   [\j_{ab},\bi_l] = ( \delta_{a,l-1} -\delta_{b,l-1}  +
    \delta_{bl} -\delta_{al})\m_{ab}} \\
\multicolumn{3}{c}{
   [\m_{ab},\bi_l] = - ( \delta_{a,l-1} -\delta_{b,l-1}  +
    \delta_{bl} -\delta_{al})\j_{ab}} \\
\end{array}
\label{ca}
\ee
\be
[\j_{ab},\m_{ab}] =-2\k_{ab}\sum_{s=a+1}^b \bi_s  \qquad\qquad
[\bi_{k},\bi_{l}]=0 .
\label{cb}
\ee
If we
discard  the condition (\ref{ag}) and consequently, further to the 
generators of $su_\k(N+1)$ we add the 
matrix $\ei=i \sum_{a=0}^{N}e_{aa}$, we obtain the {\em unitary} CK
family
$u_\k(N+1)$ of dimension  $(N+1)^2$ and with commutators  given by
(\ref{ca}), (\ref{cb}) and
\be
[\j_{ab},\ei] = 0   \qquad
[\m_{ab},\ei] = 0  \qquad
[\bi_l,\ei] = 0 .
\label{cc}
\ee
When all $\k_a\ne 0$ we find the pseudo-unitary  algebras $u(p,q)$
which are not simple (they are in the semisimple series $A_{N}\oplus
D_1$) but their pattern is similar to that of
$su(p,q)$ with respect to
$\Lam$.

The commutation rules of the {\em quaternionic unitary 
CK   family}  $\sq_\k(N+1)$ turn out to be \cite{tesis,symplectico}:
\be
\begin{array}{lll}
[\j_{ab},\j_{ac}] =  \k_{ab}\j_{bc} &\qquad
[\j_{ab},\j_{bc}] =-\j_{ac} &\qquad
[\j_{ac},\j_{bc}] = \k_{bc}\j_{ab}\cr
[\m_{ab}^\a,\m_{ac}^\a] =\k_{ab} \j_{bc} &\qquad
[\m_{ab}^\a,\m_{bc}^\a] = \j_{ac} &\qquad
[\m_{ac}^\a,\m_{bc}^\a] =\k_{bc} \j_{ab} \cr
[\j_{ab},\m_{ac}^\a] = \k_{ab}\m_{bc}^\a &\qquad
[\j_{ab},\m_{bc}^\a] =-\m_{ac}^\a &\qquad
[\j_{ac},\m_{bc}^\a] =-\k_{bc}\m_{ab}^\a \cr
[\m_{ab}^\a,\j_{ac}] =-\k_{ab}\m_{bc}^\a &\qquad
[\m_{ab}^\a,\j_{bc}] =-\m_{ac}^\a &\qquad
[\m_{ac}^\a,\j_{bc}] = \k_{bc}\m_{ab}^\a \cr
[\j_{ab},\j_{de}]= 0 &\qquad
[\m_{ab}^\a,\m_{de}^\a] =0 &\qquad
[\j_{ab},\m_{de}^\a] =0 \cr
\multicolumn{3}{l}{
  [\j_{ab},\e_d^\a] = ( \delta_{ad} -\delta_{bd})\m_{ab}^\a   
\qquad
  [\m_{ab}^\a,\e_d^\a] = -( \delta_{ad} -\delta_{bd}) \j_{ab}}\cr
\multicolumn{3}{l}{
  [\j_{ab},\m_{ab}^\a] = 2\k_{ab}(\e_{b}^\a-\e_{a}^\a)  
\qquad
  [\e_{a}^\a,\e_b^\a] = 0 }\cr
\end{array}
\label{da}
\ee
$$
\begin{array}{l}
[\m_{ab}^\a,\m_{ac}^\b] = \k_{ab}\ep_{\a\b\c}\m_{bc}^\c  \qquad
[\m_{ab}^\a,\m_{bc}^\b] = \ep_{\a\b\c}\m_{ac}^\c  \qquad
[\m_{ac}^\a,\m_{bc}^\b] = \k_{bc}\ep_{\a\b\c}\m_{ab}^\c \cr
[\m_{ab}^\a,\m_{de}^\b] = 0\qquad
[\m_{ab}^\a,\m_{ab}^\b] = 2\k_{ab}\ep_{\a\b\c}(\e_a^\c + \e_b^\c) \cr
[\m_{ab}^\a,\e_{d}^\b] =(\delta_{ad}+\delta_{bd})
\ep_{\a\b\c}\m_{ab}^\c
\qquad
[\e_{a}^\a,\e_{b}^\b] =2\delta_{ab}\ep_{\a\b\c}\e_{a}^\c    
\end{array}
$$
 where $\ep_{\a\b\c}$ is
the completely antisymmetric unit  tensor with $\ep_{123}=1$, and 
whenever   three quaternionic indices $\a$, $\b$, $\c$ 
appear, they are  assumed to be different.

Finally, we would like to remark that in addition to these three
main `signature' families of CK algebras, whose simple members
$so(p,q)$,
$su(p, q)$, $sq(p,q)$ can be realised as antihermitian matrices
over either $\Ree$, $\Cee$  or $\He$, there are other CK families
\cite{tesis,goslar}.
In the $C_{N+1}$ Cartan series, the remaining real Lie algebra is
the  symplectic $sp(2(N+1), \Ree)$, which can be
interpreted in terms of CK families either as the single simple
member of its own CK family 
$sp_{\k}(2(N+1), \Ree)$, or  alternatively, as
the unitary family
$u_{\k}((N+1), \He')$ over the algebra of  the split quaternions
$\He'$ (a pseudo-orthogonal variant of quaternions, where 
$i_1, i_2, i_3$ still anticommute, but their  squares are $i_1^2=-1,
i_2^2=1, i_3^2=1$; this is not a division algebra). 
Likewise there are other CK families associated to    the
remaining real simple Lie algebras, namely:
 $su^*(2(N+1))\approx sl(N+1, \He)$ in the Cartan series
$A_{2N+1}$, $sl(N+1,
\Ree)\approx su(N+1,\Cee')$ also in $A_N$, and
$so^*(2N)$ in  $D_N$.


\sect{Central extensions}

Let $\G$ an $r$-dimensional Lie algebra with generators
$\{X_1,\dots,X_r\}$ and structure constants
$C_{ij}^k$.  A generic central extension $\overline{\G}$ of $\G$ 
by the one-dimensional algebra generated by  $\Xi$ is a Lie
algebra with   
$(r+1)$ generators
$\{X_1,\dots,X_r,\Xi\}$ and commutation relations:
\be
[X_i,X_j]=\sum_{k=1}^r C_{ij}^k X_k  + \xi_{ij} \Xi  \qquad
[\Xi,X_i]=0 .
\label{ea}
\ee
Hence we consider an initial extension
coefficient   $\xi_{ij}$ associated to {\em each} Lie bracket
$[X_i,X_j]$.
As $\overline{\G}$ must be a Lie algebra, the  extension
coefficients  
$\xi_{ij}$  must be antisymmetric in their indices,
$\xi_{ji}=-\xi_{ij}$, and  must
fulfil the following conditions coming from the   Jacobi
identities for the generators 
$X_i, X_j, X_l$ in   $\overline{\G}$:
\be
\sum_{k=1}^r
\left(
C_{ij}^{k}\xi_{kl}+C_{jl}^{k}\xi_{ki}+C_{li}^{k}\xi_{kj}
\right) =0 .
\label{eb}
\ee
Therefore as a first step in the study of the central extensions
of a given Lie algebra $\G$ we have to solve the set
of linear equations (\ref{eb}) involving all initially possible 
extension coefficients   $\xi_{ij}$ (\ref{ea}). 
Afterwards we
have to find out  which of these coefficients are trivial, that
is, which can be removed from (\ref{ea}) by means of a
change of basis in $\overline{\G}$. Explicitly, 
if for a set of arbitrary real numbers $\mu_k$ we perform a change
of generators 
\be
X_k\to X'_k=X_k+\mu_k\Xi,
\label{ec}
\ee
then the commutators for the generators $\{X'_1,\dots,X'_r,\Xi\}$
are given by  (\ref{ea}) with a new set of extension coefficients
\be
\xi_{ij}' = \xi_{ij} -\sum_{k=1}^r C_{ij}^k \mu_k, 
\label{ed} 
\ee
where
$\delta\mu(X_i, X_j) =
\sum_{k=1}^r C_{ij}^k \mu_k$ is the two-coboundary generated by
$\mu$. Extension coefficients differing by a two-coboundary
correspond to equivalent extensions, and those extension
coefficients which are a two-coboundary, $\xi_{ij}= -\sum_{k=1}^r
C_{ij}^k \mu_k$, correspond to trivial extensions. The classes of
equivalence of non-trivial two-cocycles determine  the second
cohomology group of the Lie algebra, $H^2(\G,\Ree)$.

The procedure we have just described can be carried out for each
 family of CK algebras   in a global way, so
that this unified approach avoids at once and for all the
need of a case-by-case study for any given algebra in the family.
In general, the extension coefficients arising for
each CK family can be casted into three types according to their
behaviour under contraction (when some $\k_a$ vanish):

\noindent
$\bullet$ {\em Type I extension coefficients:} they correspond to
central extensions which are trivial for all the CK algebras
belonging to a given family simultaneously, no matter of how many
coefficients
$\k_a$ are equal to zero. Therefore they can be removed at once
by means of redefinitions as (\ref{ec}).

\noindent
$\bullet$ {\em Type II extension coefficients:} they give rise to
non-trivial extensions when some $\k_a=0$, and to trivial ones
otherwise. Therefore these extensions become non-trivial through
contractions and are called pseudoextensions
\cite{Aldaya,Azcarraga}.

\noindent
$\bullet$ {\em Type III extension coefficients:} they must fulfil
some additional conditions in such manner that whenever they are
non-zero they are always non-trivial. Therefore these extensions
cannot appear through the pseudoextension process.

In the sequel we present the complete results on the problem of
finding all non-trivial central extensions for the three main 
families of CK algebras. We will directly discard type I
extensions which are trivial for all CK algebras in each family.


\subsect{Central extensions of the orthogonal CK algebras}

 The  non-zero commutators
of any central extension $\overline{so}_{\k}(N+1)$ of the
orthogonal CK  algebra ${so}_{\k}(N+1)$ are given \cite{ortogonal} by:
\be
\begin{array}{ll}
\!\! [\om_{ab}, \om_{bc}] = - \om_{ac}    &\  \cr
\!\! [\om_{ab}, \om_{a\,b+1}] = \k_{ab} \om_{b\,b+1}+\k_{a\,b-1}
\cextiif_{b\,b+1}\Xi &\
\!\! [\om_{ab}, \om_{ac}] = \k_{ab} \om_{bc}
\quad \mbox{\rm for}\  c > b+1\cr
\!\! [\om_{ac}, \om_{a+1\,c}] = \k_{a+1\,c} \om_{a\,a+1}+\k_{a+2\,c}
 \cextiil_{a\,a+1}\Xi  &\
\!\! [\om_{ac}, \om_{bc}] = \k_{bc}\om_{ab}
\quad \mbox{\rm for}\ b > a+1  \cr
 \multicolumn{2}{l}
 {\!\! [\om_{a\,a+1}, \om_{c\,c+1}] = \cextiii_{a+1\,c+1} \Xi 
\qquad\qquad\quad  [\om_{a\,a+2}, \om_{a+1\,a+3}] = -\k_{a+2} 
  \cextiii_{a+1\,a+3} \Xi .}  
\end{array}
\label{fa}
\ee
 The extension is completely characterized by the following
extension coefficients:

\noindent
 $\bullet$   Two  type II coefficients,
$\cextiil_{01}$, 
$\cextiif_{N-1\,N}$. The extension  $\cextiil_{01}$ (resp.\ 
$\cextiif_{N-1\,N}$) is non-trivial if $\k_2=0$ (resp.\ 
$\k_{N-1}=0$) and is trivial otherwise.   

\noindent
 $\bullet$   $(N-2)$ type II  {\em pairs},
$\cextiif_{a+1\,a+2}$, $\cextiil_{a+1\,a+2}$ ($a=0,1,\dots,N-3$),
fulfilling
\be
\k_{a+3}\cextiif_{a+1\,a+2} =
\k_{a+1}\cextiil_{a+1\,a+2}.
\label{fb}
\ee
The two extensions 
$\cextiif_{a+1\,a+2}$, $\cextiil_{a+1\,a+2}$ are both non-trivial
when
$\k_{a+1} = 0$ and  $\k_{a+3} = 0$, and
both are simultaneously trivial otherwise.

\noindent
$\bullet$  $(N-1)(N-2)/2$ type III coefficients 
$\cextiii_{b+1\,d+1}$ ($b=0,1,\dots,N-3$ and $d= b+2,\dots, N-1$)
which must satisfy the following additional relations:
\be
\begin{array}{lll}
\mbox{If}\quad d=b+2:&\quad \omega \cextiii_{b+1\,b+3} = 0
  &\quad\mbox{for}\quad
\omega=\k_{b}, \k_{b+1}\k_{b+2}, \k_{b+2}\k_{b+3},  
\k_{b+4}\cr
\mbox{If}\quad d>b+2:&\quad \omega\cextiii_{b+1\,d+1}  =0
  &\quad\mbox{for}\quad
\omega=\k_{b},  \k_{b+2},  \k_{d},  \k_{d+2} 
\end{array}
\label{fc}
\ee
where it is understood that conditions as 
$\k_{0}\cextiii=0$ or $\k_{N+1}\cextiii=0$ are not
present. Whenever the extension $\cextiii_{b+1\,d+1}$ is non-zero
it is  always non-trivial.

The pseudoextension character of type II coefficients 
$\cextiif_{a+1\,a+2}$, $\cextiil_{a+1\,a+2}$ means that if 
$\k_{a+1} \ne 0$ and  $\k_{a+3} \ne 0$  both can be
simultaneously removed by applying a redefinition of the generator
$\om_{a+1\,a+2}$:
\be
 \om_{a+1\,a+2}\to  \om'_{a+1\,a+2}=
\om_{a+1\,a+2}+\frac{\cextiif_{a+1\,a+2}}{\k_{a+1}}
\Xi=\om_{a+1\,a+2}+\frac{\cextiil_{a+1\,a+2}}{\k_{a+3}}
\Xi.
\label{fd}
\ee
The equality is guaranteed by the condition (\ref{fb}). Thus in
this case both extensions are trivial. If we perform, for
instance, the contraction $\k_{a+1}\to 0$, the relation
(\ref{fb}) implies that
$\k_{a+3}\cextiif_{a+1\,a+2} = 0$, so  if $\k_{a+3}\ne 0$ the
extension $\cextiif_{a+1\,a+2}$ vanishes, while 
$\cextiil_{a+1\,a+2}$ can be eliminated by using (\ref{fd}).
Therefore both extensions become non-trivial only through the two
contractions $\k_{a+1}=0$ and $\k_{a+3}= 0$.
The same happens for the remaining  type II extensions 
$\cextiil_{01}$ and $\cextiif_{N-1\,N}$ with respect to
$\k_2$  and $\k_{N-1}$, respectively. We remark that type III
extensions do not appear under such process.

Let us illustrate now these results with some interesting
algebras \cite{ortogonal}:

\noindent
{(1)} $so(p,q)$ with $p+q=N+1$ (all $\k_a\ne 0$).\\
All type III coefficients vanish due to (\ref{fc}), and all type
II are trivial as they  can be removed by means of redefinitions
(\ref{fd}). Therefore, as it is well known (Whitehead's
lemma) these algebras  have no non-trivial extensions:
${\mbox{dim}}\,(H^2(so(p,q),\Ree))=0$.

\noindent 
{(2)} $iso(p,q)$ with $p+q=N$ ($\k_1=0$ and the others
are non-zero).\\
These   algebras include the    Euclidean $iso(N)$
and Poincar\'e $iso(N-1,1)$ ones.
The case $N=2$ is special since $\k_1=\k_{N-1}=0$, so
that  there is single non-trivial extension $\cextiif_{N-1\,N}=
\cextiif_{12}$ and ${\mbox{dim}}\,(H^2(iso(p,q),\Ree))=1$; this
extension  corresponds to a uniform and constant force field in
the $1+1$ Minkowskian free kinematics for the Poincar\'e algebra
$iso(1,1)$. However this extension is no longer possible if $N>2$
$(0=\k_1\ne
\k_{N-1})$ and there are no non-trivial type II  extensions;
furthermore  it can be checked that all type III ones vanish so that 
${\mbox{dim}}\,(H^2(iso(p,q),\Ree))=0$.

\noindent {(3)} $iiso(p,q)$  with $p+q=N-1$ ($\k_1=\k_2=0$ and the
others are non-zero).\\
We have to distinguish three cases according to the values of $N$:
\medskip

\begin{tabular}{llc}
$N$&\quad $\mbox{Non-trivial
extensions}$&\quad${\mbox{dim}}\,(H^2(iiso(p,q),\Ree))$\\ 
\hline
$2$&\quad $\cextiil_{01},\cextiif_{12}$&\quad $2$\\
$3$&\quad
$\cextiil_{01},\cextiif_{23},\beta_{13}$&\quad $3$\\
$>3$&\quad  $\cextiil_{01}$ &\quad $1$\\
\hline
\end{tabular}

\medskip

\noindent
The  physical role of these extensions for
the Galilean algebra 
$iiso(N-1)$  is as follows: the only non-trivial coefficient which
appears for any
$N$,
$\cextiil_{01}$, is  the mass; 
$\cextiif_{12}$ is a constant force for $iiso(1)$;  
$\cextiif_{23}$ is a sort of non-relativistic remainder of the
Thomas precession  for  
  $iiso(2)$ \cite{balls} and $\beta_{13}$ has no physical
meaning  and disappears once we move from the Galilean 
algebra to the corresponding Lie group.

\noindent {(4)}  $ii\ldots iso(1)$ (all $\k_a=0$).\\
This is the most
contracted CK algebra: the orthogonal flag algebra. All the
conditions (\ref{fb}) and (\ref{fc}) are fulfilled and all
possible extensions are non-trivial, that is, there are $2(N-1)$
type II and $(N-1)(N-2)/2$ type III non-trivial extensions.


\subsect{Central extensions of the unitary CK algebras}

The  commutation relations of any central extension
$\overline{su}_\k(N+1)$ of the special unitary CK algebra
${su}_\k(N+1)$ can be written \cite{unitario} as the commutation
relations  (\ref{ca}) together with:
\be
 \displaystyle  [\j_{ab},\m_{ab}] = -2\k_{ab}\sum_{s=a+1}^b \bi_s
+\sum_{s=a+1}^b\k_{a\,s-1}\k_{sb}\,\cexta_{s} \Xi  \qquad
[\bi_k,\bi_l]=\cextb_{kl}\Xi\quad k<l
\label{ga}
\ee
which will replace those in (\ref{cb}). The possible extension
coefficients are

\noindent
$\bullet$ $N$ type II coefficients ${\cexta_{k}}$
($k=1,\dots,N$). Each of them gives rise to a non-trivial
extension if
$\k_k= 0$ and to a trivial one otherwise.

\noindent
$\bullet$ $N(N-1)/2$  type III coefficients $\cextb_{kl}$
($k<l$ and
$k,l=1,\dots,N$), satisfying
\be
 \k_k \cextb_{kl}=0\qquad \k_l\cextb_{kl}=0.
\label{gb}
\ee
Thus, $\cextb_{kl}$ vanishes when at least
one of the parameters $\k_k,\ \k_l$ is different from zero. When
$\cextb_{kl}$ is non-zero  it  is always non-trivial.

On the other hand, the  Lie brackets of any central extension
$\overline{u}_\k(N+1)$ of the unitary CK algebra
${u}_\k(N+1)$ are given \cite{unitario} by
(\ref{ca}), (\ref{ga}) and:
\be
\begin{array}{lll}
[\j_{ab},\ei] = 0  & \qquad
[\m_{ab},\ei] = 0  & \qquad
[\bi_k,\ei]=\cextg_{k}\Xi
\label{gc}
\end{array}
\ee
which will replace those in (\ref{cc}). 
The
extension is completely characterized by  the
above extension coefficients ${\cexta_{k}}$ and $\cextb_{kl}$
together with:

\noindent
$\bullet$
$N$ type III coefficients ${\cextg_{k}}$ ($k=1,\dots,N$)
satisfying
\be
 \k_k \cextg_{k}=0.
\label{gd}
\ee
When $\cextg_k$ is non-zero, the extension that it determines  is
non-trivial.

Unlike the  orthogonal CK family we can give now the following
closed expressions for the dimension of the second cohomology
group of these two unitary CK families which establish the
number of non-trivial extension coefficients according to the
number  $n$  of coefficients $\k_k$ which are equal to zero for a
given unitary CK algebra:
\be
{\mbox{dim}}\,
(H^2({su}_\k(N+1),\Ree))=n + \frac {n(n-1)}{2} =
  \frac{n(n+1)}2
\label{gh}
\ee
\be
{\mbox{dim}}\,
(H^2({u}_\k(N+1),\Ree))=n + \frac {n(n-1)}{2} +n =
  \frac {n(n+3)}2.
\label{gi}
\ee
The first term $n$ in the sum of (\ref{gh}) and (\ref{gi})
corresponds to the central extensions ${\cexta_{k}}$, the second
term $\frac {n(n-1)}{2}$ to the $\cextb_{kl}$ and the third term
$n$ in (\ref{gi}) to the  $\cextg_k$.

As far as the special unitary CK algebras is concerned, we find
that  the second cohomology group is trivial for the
simple $su({p,q})$ algebras with all
$\k_a\ne 0$ ($n=0$). 
The inhomogeneous
$iu({p,q})$ algebras with 
$\k_1=0$ and all other constants $\k_a
\neq 0$ ($n=1$) have, in any dimension, a single non-trivial
extension: $\cexta_{1}$. The algebras with $\k_1=\k_2=0$ and the
remaining $\k_a \neq 0$ ($n=2$) have always three 
non-trivial
extensions: $\cexta_{1}$, $\cexta_{2}$ and $\beta_{12}$.
The  special unitary flag algebra with all $\k_a= 0$ ($n=N$) has
the maximum number of non-trivial extensions: ${N(N+1)}/2$.
A similar discussion can be performed for the CK family 
${u}_\k(N+1)$.


\subsect{Central extensions of the quaternionic unitary CK
algebras}

The algebraic structure of the CK family
${sq}_\k(N+1)$ seems to be much more complicated
than those in the orthogonal and unitary CK families. However, the
solution to the central extension problem comes as a surprise and
is quite simple to state \cite{symplectico}:
all  central extensions of any Lie algebra belonging to the
quaternionic unitary CK family ${sq}_\k(N+1)$ are always trivial
for any $N$ and for any  values of the set of
constants $\k_1, \k_2, \dots, \k_N$:
\be
{\mbox{dim}}\,
(H^2({sq}_\k(N+1),\Ree))=0 .
\label{ha}
\ee


{\section*{Acknowledgments}} 
F.J.H. would like to acknowledge the organisers of the III 
IWCQIS for their invitation and hospitality during the meeting at
Yerevan. This  work was partially supported by DGICYT  (project 
PB94--1115) from the Ministerio de Educaci\'on y Cultura de
Espa\~na  and by Junta de Castilla y Le\'on (Projects CO1/396 and
CO2/297).


\end{document}